\documentstyle[12pt]{article}

\begin{document}
\newcommand{\an}{\alpha_s^{naive}}
\newcommand{\aex}{\alpha_s^{exact}}
\thispagestyle{empty}
{\hfill Preprint INR-925/96}

\hfill{July 1996}

\vspace*{2cm}
\begin{center}

{\Large \bf Infrared modified analysis for the $\tau$ lepton width}
\vskip 1.2cm
N.V. Krasnikov and A.A. Pivovarov

{\it Institute for Nuclear Research of the Russian Academy of Sciences,
Moscow 117312}

\vskip 1.2cm

\end{center}
\centerline{\bf Abstract  }

\noindent
We argue that accounting for higher order corrections in QCD
by integrating a running coupling constant through an infrared region 
can be most easily done 
with making use of a scheme without the Landau pole.
Within this approach the entire ambiguity of the answer is identical 
to that of the choice of renormalization scheme.
The uncertainties for the $\tau$
lepton width resulting from such a technique are discussed.

\vskip 5cm

\begin{center}

Talk given at 10th International Conference

                PROBLEMS OF QUANTUM FIELD THEORY

           Alushta (Crimea, Ukraine)  13 - 18 May 1996        
\end{center}

\newpage
This talk is based essentially on our recent paper \cite{krpiv0}
and gives a brief description of our suggestion to treat 
those higher order corrections of perturbation theory that stem 
from the integration of a running coupling constant over an infrared 
region. 
Formulated in a slightly different context the problem of accounting
for such corrections is known as a problem of infrared renormalons 
\cite{allren}
and of proper determination of their contribution to physical
observables.
We propose the way of definition of these ambiguous contributions
based on a continuation from perturbation theory region by using the
freedom of the renormalization scheme choice. We parametrize this
freedom with a special definition of the $\beta$ function that is
considered to be valid in the whole region of the copling constant
(even in the regime of strong coupling).
Then we introduce a particular function 
that gives no singularity for the corresponding
running coupling constant.

Our main attention is concentrated on the analysis of the $\tau$
lepton decay width that is represented by integrals over
the infrared region. Several ways to define them using the
freedom of choice of the RG scheme are given
\cite{krpiv0}. The main conclusion we draw is
that the results of integration 
can easily be made well defined without any
explicit nonperturbative contributions.
These results are ambiguous 
to the same degree as any ordinary PT series,
numerically it can be important because in the infrared region the
coupling constant becomes large in most of ``natural'' RG schemes.
However it can be made small as well 
by some particular choice of extrapolation to low momenta.

The problem of the $\tau$ lepton width has been 
widely discussed in the literature 
(as some recent references see, e.g. \cite{alt})
so we limit ourselves to qualitatively different versions
of changing RG schemes only. 
We propose a set of schemes that regularize the infrared
behavior of the coupling constant 
and allow one to use any reference scheme for
high energy domain.
All these schemes are legal and perturbatively equivalent at high
energies. Nevertheless numerical 
uncertainties that come from low energy region
are quite essential.

We introduce our main object -- a $\kappa$ scheme that is
determined by the following $\beta$ function \cite{krpiv0} 
\begin{equation}
\beta_\kappa(a)={ \beta(a)\over 1-\kappa a^n \beta(a)}
\label{kappa}
\end{equation}
where $\beta(a)$ is a $\beta$ function in a reference scheme,
$\overline{\rm MS}$ 
for instance. 
The form (\ref{kappa})
is chosen because of practical
convenience only -- it requires no more technical work for obtaining
practical results than a corresponding
reference scheme and eliminates the Landau pole in infrared region.

The $\beta$ function given by eq.~(\ref{kappa}) is bounded at large
$a$ and the RG equation has a solution for $a(z)$ that is
defined on the whole positive semiaxis and is free of the Landau
pole.
The absence of singularities (Landau ghost) 
allows one to use the evolution 
of the coupling constant till the very origin.
The solution to the RG equation for the invariant charge $a(z)$
in the
$\kappa$ scheme is simple because it is closely related
to the $\overline{\rm MS}$
running coupling constant
\[
ln(s/\Lambda^2)=\Phi(a)-\kappa {a^{n+1}\over n+1},
\]
\begin{equation}
\Phi(a)={1\over a}-c\ln({1\over a}+c)
+\int_0^a\left({1\over \beta(\xi)}
-{1\over \beta_{(2)}(\xi)}\right)d\xi
\label{phinorm}
\end{equation}
with $\beta_{(2)}(a)=-a^2(1+ca)$ and with
the standard definition of the parameter $\Lambda$
to be $\Lambda_{\overline{\rm MS}}$.
Two charges in $\kappa$ and $\overline{\rm MS}$ are connected through
$$
a_{\kappa}=a-{\kappa\over n+1}a^{n+3} + o(a^{n+3}).
$$ 
The exact connection can be found from eq.~(\ref{phinorm}).
So, for $n>1$ they coincide almost for all observables because in
practice there are no calculations beyond the third order.

We now consider the uncertainties for 
predictions of the $\tau$ lepton
width or for the parameter $\alpha_s(m_\tau^2)$ extracted from
experimental data on this width.
The expression for the $\tau$ lepton width
has the form \cite{bra} 
\begin{equation}
R_\tau=\int_0^{m_\tau^2}{ds\over m_\tau^2}2(1-{s\over
m_\tau^2})^2(1+2{s\over m_\tau^2})R(s)
\label{tautheor}
\end{equation}
with $R(s)$ given by 
$
R(s)=3(1+{\alpha_s(s)\over \pi}+\ldots).
$
Using the experimental value
$
R_\tau^{exp}=3.645\pm0.024
$
we find after the standard analysis in $\overline{\rm MS}$ scheme
$
\alpha_s(m_\tau^2)=0.353\pm 0.008.
$

Now we give results in the $\kappa$ scheme
that is more general and allows the integration
over infrared region.
This is a generalization of the fixed point approach.
Note that the analytic continuation from Euclidean region 
\cite{anal,tau} is a
particular case of such kind an analysis.
The spectral density in $\kappa$ scheme 
(with $n=0$) up to third order is
\[
\rho(s)=a_\kappa+k_1a_\kappa^2+(k_2+\kappa)a_\kappa^3,
\quad
a_\kappa=a-\kappa a^3,\quad
a=a_\kappa+\kappa a_\kappa^3,
\]
$k_1=0.7288$, $k_2=-2.0314$ \cite{gorishn,levan}.
Here the normalization of the coupling constant is chosen to be
$a=\beta_0 \alpha_s$
and only the proper hadronic part of the entire spectral density is
introduced (for more details see \cite{krpiv0}).
The expansion of $\beta$ function reads
$$
\beta(a_\kappa)
=-a_\kappa^2(1+c a_\kappa+(c_1-\kappa) a_\kappa^2+\ldots)
$$
$c=64/81$, $c_1=3863/4374$ \cite{befunc}. Thus, introducing the 
$\kappa$ scheme is equivalent at high energies to a change of $c_1$.
At low energies however two charges are different.

Integration for moments can be easily rewritten in terms of the charge
itself
$$
r_N=\int_0^{m_\tau^2}{ds\over m_\tau^2}
({s\over m_\tau^2})^N\rho(a)
=\int_\infty^{a_\kappa}
exp(N+1)(\Phi(\xi)-\kappa\xi-\Phi(a_\kappa)+\kappa a_\kappa)
({1\over \beta(\xi)}-\kappa)\rho(\xi)d\xi.
$$
Introducing the variable $\zeta=1/\xi$ we get the practical version
$$
r_N=\int_0^{a_\kappa^{-1}}
exp(N+1)(\Phi(\zeta^{-1})-\kappa\zeta^{-1}-\Phi(a_\kappa)+\kappa a_\kappa)
\left({1\over \zeta^2 \beta(\zeta^{-1})}+\kappa \zeta^2\right)
\rho(\zeta^{-1})d\zeta.
$$

Results depend on $\kappa$. This is the ordinary RG dependence
that is strong enough because the accuracy is different at large
momenta where we keep only the expansion and at small momenta where the
exact formulae have to be used to make integrals finite.
The obtained results are given in Table \ref{t:1}.
Here $a_\kappa$ 
is found from integration, $\an$ is found from the naive
(to third order) relation between the schemes
$$
\an={4\pi\over 9}(a_\kappa+\kappa a_\kappa^3),
$$
while $\aex$ is found from exact formulae for RG scheme relations  
(\ref{phinorm}).
For finding the parameter $a_\kappa$ from the integral it is
useful to know the derivative of the integral with respect to
the boundary value $a_\kappa(m_\tau^2)\equiv a_0$
$$
{dr\over da_0}=-{1\over \beta_\kappa(a_0)}2(r_0-r_2+r_3/2).
$$

\begin{table}[t]
\begin{center}
\begin{tabular}{|c|c|c|c|} \hline
$\kappa$ &$a_\kappa$ &$\an$       &$\aex$     \\ \hline
1.5      &0.2425      &0.368(16)   &0.380(18)  \\ 
1.6      &0.2209      &0.333(09)   &0.342(10)  \\ 
1.7      &0.2118      &0.318(08)   &0.326(09)  \\ 
1.8      &0.2068      &0.311(07)   &0.319(08)  \\ 
1.9      &0.2037      &0.307(07)   &0.315(08)  \\ 
2.0      &0.2016      &0.304(07)   &0.313(08)  \\ 
2.1      &0.2003      &0.303	   &0.312	\\
2.2      &0.1993      &0.303	   &0.311	\\ 
2.3      &0.1986      &0.303	   &0.312	\\ 
2.4      &0.1982      &0.303	   &0.312	\\ 
2.5      &0.1979      &0.303	   &0.313	\\ 
2.6      &0.1977      &0.304	   &0.315	\\ 
2.7      &0.1975      &0.305	   &0.316	\\ 
2.8      &0.1975      &0.306	   &0.318	\\ 
2.9      &0.1975      &0.307	   &0.319	\\ 
3.0      &0.1975      &0.308	   &0.321	\\ 
3.2      &0.1976      &0.310	   &0.325 	\\ 
3.4      &0.1978      &0.313	   &0.329	\\ 
3.6      &0.1980      &0.315	   &0.334	\\ \hline
\end{tabular}
\caption{$\kappa$ scheme results}
\label{t:1}
\end{center}
\end{table}

One can find the derivative RG equation for the coupling constant $a_\kappa$
describing its dependence on the parameter $\kappa$
\begin{equation}
{da_\kappa\over d\kappa}=a_\kappa\beta_\kappa(a_\kappa).
\label{kapeq}
\end{equation}
This is a particular case of RG equations
$$ 
{da\over dc_n}
=-\beta(a)\int_0^a{x^{n+3}dx\over \beta^2(x)}, \quad n\ge 1,
$$
that describe the dependence of the running coupling constant
on coefficients of the $\beta$ function.
Note that the dependence on $c$ is fixed by the choice of the 
parameter $\Lambda$ to be $\Lambda_{\overline{\rm MS}}$.
The extraction of $a_\kappa$ is done under the assumption of RG
invariance 
of the expression for the $\tau$ lepton width
so the only reliable data for $a_\kappa$
can be taken from that part of Table 1 where
equation (\ref{kapeq}) is satisfied. This equation can be easily solved
analytically 
(it is a linear equation if $\kappa$ is considered as a
dependent function and $a_\kappa$ as an independent variable), 
we have preferred however to solve it numerically in the vicinity of
the value ($\kappa=2.1, a_\kappa=0.2003$). 
The solution does not match well
the pattern of the dependence of the extracted $a_\kappa$ presented in
Table 1 that
means that higher order terms of perturbative expansion 
for the width are essential.
 
Note that contrary to possible impression 
the prediction in fixed point scheme, or in K scheme (see \cite{krpiv0}),
is also
nonstable. Indeed, it is easy to introduce a set of schemes
parameterized with the fixed point value of the invariant charge --
the extracted $\alpha_s(m_\tau^2)$ will depend on the scheme within
the set.
A $\beta$ function for such a set could have the form
\begin{equation}
\beta_f(a)=\beta(a)(1+\kappa\beta(a))
\label{betafixpnt}
\end{equation}
that introduces a dependence on an external scheme parameter $\kappa$.
All schemes of the type (\ref{betafixpnt})
have a fixed point with different value of the coupling
constant depending on the parameter $\kappa$.   
  
In fact, the dependence of the extracted values 
of the coupling constant for the $\tau$ lepton width on 
schemes is rather large because the energy scale ($m_\tau^2$)
is quite low. So, it might be reasonable to fix the
scheme in an arbitrary, and somehow simplest, way 
and then to parameterize the
low momenta region in the integral sense only without
detailed description of the behavior of the running
coupling from point to point. This can be done in terms of
distribution. Adding the localized distribution
like $\delta$ function and its derivatives one can make
the integral (\ref{tautheor}) well defined \cite{ren1,bub}. 
These localized contributions look
as nonperturbative terms.
Other phenomenological applications, e.g.
\cite{big}, have been also considered.

To conclude, the integration over an infrared region involves the strong
coupling dynamics and is arbitrary to a large extent.
In case of the $\tau$ lepton width the uncertainties for extracted
numerical value of the coupling constant are quite large.


\end{document}